\documentclass[10pt,a4paper,twoside]{article} 
\usepackage[T1,T2A]{fontenc}
\usepackage[utf8]{inputenc}

\newcommand{\issueYear}{0000}
\newcommand{\issueVolume}{0}

\usepackage[english,russian]{babel}
\usepackage{amsmath,amsfonts,amssymb,amscd,euscript}
\usepackage{mathrsfs}
\usepackage{epsfig,epstopdf}
\usepackage{manyfoot}

\usepackage{anyfontsize}

\usepackage[%
	pdfpagemode=UseNone,%
    pdfpagelayout=TwoPageLeft,%
    bookmarks=false,%
    bookmarksopenlevel=1,%
	breaklinks,%
]{hyperref}        
\usepackage[all]{hypcap}
\usepackage{cite}

\addto\captionsrussian{%
  \renewcommand{\UDKName}{УДК}%
  \renewcommand{\KeyWords}{Ключевые слова}%
  \renewcommand\refname{\normalsize Список литературы}%
  \renewcommand{\citeString}{
    \textbf{Просьба ссылаться на эту статью следующим образом:}\\
    {\the\authorslistInv} {\csname title\endcsname}. {\it Пространство, время и фундаментальные взаимодействия}. \issueYear. №~\issueVolume. \mbox{C.~\pageref{\theArticle:article:fstpage}–-\pageref{\theArticle:article:lastpage}}.
  }%
  \renewcommand{\issueMnths}{\issueMnthsRu}%
}
\addto\captionsenglish{%
  \renewcommand{\UDKName}{UDC}%
  \renewcommand{\KeyWords}{Keywords}%
  \renewcommand\refname{\normalsize References}%
  \renewcommand{\citeString}{
    \textbf{Please cite this article in English as:}\\
    {\the\authorslistInvSub} {\csname titleSub\endcsname}. {\it Space, Time and Fundamental Interactions}, \issueYear, no.~\issueVolume, \mbox{pp.~\pageref{\theArticle:article:fstpage}--\pageref{\theArticle:article:lastpage}}.
  }%
  \renewcommand{\issueMnths}{\issueMnthsEn}%
}

\addto\extrasrussian{}
\addto\extrasenglish{}

\renewcommand{\footnotesize}{\fontsize{8}{10}\selectfont}
\renewcommand{\small}{\fontsize{9}{12}\selectfont}
\renewcommand{\normalsize}{\fontsize{10}{14}\selectfont}

\usepackage[scaled=1.2]{PTSansNarrow}
\usepackage[scaled=.9]{PTMono}
\newcounter{Article}
\renewcommand{\theArticle}{\arabic{Article}}
\numberwithin{equation}{section}

\usepackage{xstring}

\usepackage[explicit]{titlesec}

\titleformat{\section}%
  {\normalfont\bfseries}%
  {}%
  {0em}%
  {\normalfont\bfseries\textbf{\thesection.\hspace{0.02em} #1}}
  
\titleformat{name=\section,numberless}%
  {\normalfont\bfseries}%
  {}%
  {0em}%
  {\normalfont\bfseries #1}

\titlespacing*{\section}{0pt}{1em}{1em}

\titleformat{\subsection}%
  {\normalfont\itshape\bfseries}%
  {}%
  {0em}%
	{\normalfont\bfseries\thesubsection.\hspace{0.2em} #1}
  
\titleformat{name=\subsection,numberless}%
  {\normalfont\itshape}%
  {}%
  {0em}%
  {\normalfont\itshape #1}

\titlespacing*{\subsection}{0pt}{1em}{1em}

\renewcommand*\thesection{\arabic{section}}

\newcommand{\DOIM}{DOI: 10.17238/issn2226-8812.2018.2}

\usepackage{caption}
\DeclareCaptionFont{white}{\color{white}}
\DeclareCaptionFormat{overlay}{\small{\bfseries #1.}{\hskip 1em}#3}
\DeclareCaptionFormat{empty}{}
\usepackage{fancyhdr}
\fancypagestyle{firstpage}
{
  \fancyhf{}

	\fancyhead[LE,LO]{\footnotesize ПРОСТРАНСТВО, ВРЕМЯ И ФУНДАМЕНТАЛЬНЫЕ ВЗАИМОДЕЙСТВИЯ}
  \fancyhead[RE,RO]{\footnotesize \issueYear, Вып. \issueVolume}
}

\fancypagestyle{firstpage_arxiv}
{
  \fancyhf{}

	\fancyhead[LE,LO]{\footnotesize SPACE, TIME AND FUNDAMENTAL INTERACTIONS}
  \fancyhead[RE,RO]{\footnotesize \issueYear, vol. \issueVolume}
}

\fancypagestyle{outputpage}{
  \fancyhf{}

  \fancyhead[LE,LO]{\DOIM}
  \fancyhead[RE,RO]{\issueMnths\quad \issueYear}
}
\fancypagestyle{commonpage}{
  \fancyhf{}

  \fancyhead[LE,RO]{\normalsize\thepage}
  \fancyhead[LO]{\footnotesize\rightmark}
  \fancyhead[RE]{\footnotesize\leftmark}
}
\renewcommand{\sectionmark}[1]{}
\renewcommand{\subsectionmark}[1]{}
\makeatletter
\g@addto@macro\bfseries{\boldmath}
\makeatother
\makeatletter
\let\@afterindentfalse\@afterindenttrue
\makeatother

\usepackage{listings}
\lstdefinelanguage{Maple}
{
  morekeywords={
    abs,animate,array,assuming,
    Christoffel1,Christoffel2,collect,create,cov_diff,
	d1metric,d2metric,DEplot,diff,Dirac,display,do,dsolve,dual,
    Einstein,end,eval,evalf,expand,
	for,from,
	Heaviside,
	if,implicitplot,innerprod,int,interface,invert,
	geodesic_eqns,get_char,get_compts,get_rank,grad,
	Levi_Civita,lin_com,local,lhs,
	matadd,
	odeplot,op,
	piecewise,plot,plot3d,point,pointplot,proc,prod,
	Ricci,Ricciscalar,Riemann,restart,rhs,
	scalarmul,seq,signum,simplify,solve,spacecurve,sqrt,subs,sum,
	to,trunc,
	union,
	with
  },
  sensitive=false, 					
  morecomment=[l]{\#}, 				
  morecomment=[s]{/*}{*/}, 			
  morestring=[b]" 					
}
\newtoks{\affillistRu}
\newtoks{\affillistEn}
\newtoks{\mailList}
\newtoks{\authorslist}
\newtoks{\authorslistSub}
\newtoks{\authorslistInv}
\newtoks{\authorslistInvIndex}
\newtoks{\authorslistInvIndexSub}
\newtoks{\authorslistFooter}
\newtoks{\authorslistFooterSub}
\newtoks{\authorslistInvSub}
\newcount\articleslanguage
\newcount\loopingindex 

\newcommand{\KeyWords}{}
\newcommand{\UDKName}{}

\newcommand{\citeString}{}
\newcommand{\issueMnths}{}

\newlength\boxheight
\newcounter{numauthors}
\newcommand{\affili}[3]{
  \global\affillistRu=\expandafter{\the\affillistRu \text{$^{#1 \ }$} {#2}  \newline}
	\global\affillistEn=\expandafter{\the\affillistEn \text{$^{#1 \ }$} {#3}  \newline}
}
\newcommand{\addauthor}[1]{%
  \global\authorslist=\expandafter{\the\authorslist#1}
}
\newcommand{\addauthorSub}[1]{%
  \global\authorslistSub=\expandafter{\the\authorslistSub#1}
}
\newcommand{\addauthorInv}[1]{%
  \global\authorslistInv=\expandafter{\the\authorslistInv#1}
}
\newcommand{\addauthorInvIndex}[1]{%
  \global\authorslistInvIndex=\expandafter{\the\authorslistInvIndex#1}
}
\newcommand{\addauthorInvIndexSub}[1]{%
  \global\authorslistInvIndexSub=\expandafter{\the\authorslistInvIndexSub#1}
}
\newcommand{\addauthorFooter}[2]{%
  \global\authorslistFooter=\expandafter{\the\authorslistFooter\vspace{10pt}\\ \noindent#1\\ E-mail: #2}
}
\newcommand{\addauthorFooterSub}[2]{%
  \global\authorslistFooterSub=\expandafter{\the\authorslistFooterSub\vspace{10pt}\\ \noindent#1\\ E-mail: #2}
}
\newcommand{\addauthorInvSub}[1]{%
  \global\authorslistInvSub=\expandafter{\the\authorslistInvSub#1}
}
\newcommand{\DOI}[1]{\expandafter\gdef\csname doi\endcsname{#1}}
\newcommand{\UDK}[1]{%
\expandafter\gdef\csname udk\endcsname{#1}
\refstepcounter{Article} 
\setcounter{numauthors}{0}
\global\authorslistSub={}
\global\authorslistInv={}
\global\authorslistInvIndex={}
\global\authorslistInvIndexSub={}
\global\authorslistInvSub={}
\global\authorslist={}
\global\affillistRu={}
\global\affillistEn={}
\global\mailList={}
\global\authorslistFooter={}
\global\authorslistFooterSub={}
}

\newcommand{\PACS}[1]{\expandafter\gdef\csname pacs\endcsname{#1}}
\newcommand{\Grant}[1]{\expandafter\gdef\csname grant\endcsname{\Footnote{\,*}{#1}}}

\newcommand{\Author}[7]{%
	\refstepcounter{numauthors}
	\expandafter\gdef\csname author\thenumauthors:name\endcsname{#1}
	\expandafter\gdef\csname author\thenumauthors:contactinfo\endcsname{#2}
	\expandafter\gdef\csname author\thenumauthors:mail\endcsname{\href{mailto:#3}{\MakeLowercase{\texttt{#3}}}}
	\expandafter\gdef\csname author\thenumauthors:nameSub\endcsname{#4}
	\expandafter\gdef\csname author\thenumauthors:contactinfoSub\endcsname{#5}
	\expandafter\gdef\csname author\thenumauthors:affill\endcsname{#6}
	\global\mailList=\expandafter{\the\mailList \footnotetext[#7]{E-mail: #3}}
	\expandafter\gdef\csname author\thenumauthors:numaut\endcsname{\,#7}
	\expandafter\gdef\csname author\thenumauthors:numautor\endcsname{#7}
	\addauthorFooter{#2}{#3}%
	\addauthorFooterSub{#5}{#3}%
	\ifnum\thenumauthors=1\addauthor{#1}\else\addauthor{,\space#1}\fi%
	\ifnum\thenumauthors=1\addauthorSub{#4}\else\addauthorSub{,\space#4}\fi%
	\ifnum\thenumauthors=1\addauthorInv{\invertName{#1}}\else\addauthorInv{,\space\invertName{#1}}\fi%
	\ifnum\thenumauthors=1\addauthorInvSub{\invertName{#4}}\else\addauthorInvSub{,\space\invertName{#4}}\fi%
	\ifnum\thenumauthors=1\addauthorInvIndex{   \invertName{#1}$^{#6,}$\footnotemark[#7]}\else   \addauthorInvIndex{,\space\invertName{#1}$^{#6,}$\footnotemark[#7]}\fi
	\ifnum\thenumauthors=1\addauthorInvIndexSub{\invertName{#4}$^{#6,}$\footnotemark[#7]}\else\addauthorInvIndexSub{,\space\invertName{#4}$^{#6,}$\footnotemark[#7]}\fi
	\expandafter\gdef\csname author\thenumauthors:invname\endcsname{\invertName{#1}}
	\expandafter\gdef\csname author\thenumauthors:invnameSub\endcsname{\invertName{#4}}
	\expandafter\gdef\csname author\thenumauthors:invnameSubCaps\endcsname{\invertNameCaps{#4}}
	}

\newcommand{\Title}[3]{
\expandafter\gdef\csname titleShort\endcsname{#1}
\expandafter\gdef\csname title\endcsname{#2}
\expandafter\gdef\csname titleSub\endcsname{#3}
}

\newcommand{\Abstract}[2]{%
\expandafter\gdef\csname abstract\endcsname{#1}
\expandafter\gdef\csname abstractSub\endcsname{#2}
}
\newcommand{\Key}[2]{%
\expandafter\gdef\csname key\endcsname{#1}
\expandafter\gdef\csname keySub\endcsname{#2}
}
\newcommand{\Datereceive}[1]{%
\expandafter\gdef\csname datereceive\endcsname{#1}
}
\newcommand\Header{%
	\setcounter{equation}{0}
	\setcounter{enumiv}{0}
	\setcounter{figure}{0}
	\setcounter{table}{0}
	\setcounter{footnote}{0}
	\setcounter{section}{0}

  \thispagestyle{firstpage}
  \label{\theArticle:article:fstpage}
	\begin{flushleft} 
		\hbox{\UDKName \ \csname udk\endcsname} \par\vspace{5pt} 
		\copyright \,{} \the\authorslistInv,   \issueYear \par	\vspace{20pt} 
		\begingroup
			\textbf{{\expandafter\MakeUppercase{\csname title\endcsname}}{\csname grant\endcsname}}
		\endgroup
	\end{flushleft}
	\markboth{\the\authorslist}{\the\authorslist} 
	\markright{\csname titleShort\endcsname} 
	\the\authorslistInvIndex 
	\the\mailList \par\vspace{5pt}
	\noindent{\the\affillistRu} \par \vspace{-3pt} 
	\noindent{\small \csname abstract\endcsname} \par\vspace{8pt} 
	\noindent{\small \textit{\KeyWords}: \csname key\endcsname.} 
	\par\vspace{10pt}
}

\newcommand\SubHeader{ \normalsize%
	\begin{flushleft}	
		\begingroup
			{\noindent\bf \expandafter\MakeUppercase\csname titleSub\endcsname}
		\endgroup  
	\end{flushleft}
	\the\authorslistInvIndexSub \par\vspace{5pt}
	\noindent\the\affillistEn \par\vspace{-3pt}	
	\noindent{\small\csname abstractSub\endcsname} \par\vspace{8pt} 
	\noindent{\small \textit{\KeyWords}: \csname keySub\endcsname.} \par\vspace{10pt}
	\noindent{PACS: \csname pacs\endcsname} \par 
	\normalsize \noindent{DOI: \csname doi\endcsname}\par\vspace{20pt} 
} 

\newcommand\Footer{
	{\normalsize\vspace{8pt}\noindent 
	\textbf{Авторы}
		\the\authorslistFooter}\\[1em]
  \citeString\label{\theArticle:article:lastpage}
}

\newcommand\FooterSub{%
  \vspace{20pt}\noindent 
  {\text{\textbf{Authors}}\vspace{-5pt}
		\the\authorslistFooterSub}\\[1em]
  \citeString\label{\theArticle:article:lastpage}
}

\newcommand\SubHeaderArxiv{%
  \captionsenglish

	\setcounter{equation}{0}
	\setcounter{enumiv}{0}
	\setcounter{figure}{0}
	\setcounter{table}{0}
	\setcounter{footnote}{0}
	\thispagestyle{firstpage_arxiv}
  \label{\theArticle:article:fstpage}
	\begin{flushleft} 
		\hbox{\UDKName \ \csname udk\endcsname} \par\vspace{5pt} 
		\copyright \,{} \the\authorslistInvSub, \issueYear \par	\vspace{20pt} 
		\begingroup
			\textbf{{\expandafter\MakeUppercase{\csname titleSub\endcsname}}{\csname grant\endcsname}}
		\endgroup
	\end{flushleft}
	\markboth{\the\authorslistSub}{\the\authorslistSub} 
	\markright{\csname titleShort\endcsname} 
	\the\authorslistInvIndexSub 
	\the\mailList \par\vspace{5pt}
	\noindent{\the\affillistEn} \par \vspace{-3pt} 
	\noindent{\small \csname abstractSub\endcsname} \par\vspace{8pt} 
	\noindent{\small \textit{\KeyWords}: \csname keySub\endcsname.} 
	\par\vspace{10pt}
	\noindent{PACS: \csname pacs\endcsname} \par 
	\normalsize \noindent{DOI: \csname doi\endcsname}\par\vspace{20pt} 
}

\makeatletter
\def\@biblabel#1{#1.}
\renewenvironment{thebibliography}[1]     
	 {\vspace{1em}
	 \section*{\refname}%
      \list{\@biblabel{\@arabic\c@enumiv}}%
           {\setlength{\labelwidth}{0pt}
            \setlength{\labelsep}{.5em}
            \setlength{\leftmargin}{0pt}
			\advance\itemindent\labelsep
            \small
            \@openbib@code
            \usecounter{enumiv}%
            \let\p@enumiv\@empty
            \renewcommand\theenumiv{\@arabic\c@enumiv}}%
      \sloppy
	  \setlength{\itemsep}{-.3ex}%
      \clubpenalty4000
      \@clubpenalty \clubpenalty
      \widowpenalty4000%
      \sfcode`\.\@m}
     {\def\@noitemerr
       {\@latex@warning{Empty `thebibliography' environment}}%
      \endlist}
\makeatother

\makeatletter
\newcommand{\invertNameCaps}[1]{%
	\let\my@initials\@empty%
    \let\my@surname\@empty%
    \expandafter\@parsse\MakeUppercase{#1}~\@nil 
    \my@surname~\my@initials
}
\def\@parsse#1~#2\@nil{
  \def\argg@two{#2}%
    \ifx\argg@two\@empty%
		\edef\my@surname{#1}%
    \else%
        \edef\my@initials{\if\my@initials\@empty\else\my@initials.\fi#1}%
        \expandafter\@parsse#2\@nil%
    \fi
}

\makeatletter
\newcommand{\invertName}[1]{%
	\let\my@initials\@empty%
    \let\my@surname\@empty%
    \expandafter\@parse#1~\@nil 
    \my@surname~\my@initials
}
\def\@parse#1~#2\@nil{
  \def\arg@two{#2}%
    \ifx\arg@two\@empty%
		\edef\my@surname{#1}%
    \else%
        \edef\my@initials{\if\my@initials\@empty\else\my@initials.\fi#1}%
        \expandafter\@parse#2\@nil%
    \fi
}

\makeatletter
\def\ifemptyarg#1{%
  \if\relax\detokenize{#1}\relax 
    \expandafter\@firstoftwo
  \else
    \expandafter\@secondoftwo
  \fi}

\usepackage{graphicx}
\usepackage{dcolumn}
\usepackage{tabularx}

\begin{document}
\UDK{530.122, 524.74} 

\Title
	{Orbits of massive particles in a spherically symmetric gravitational field in view of cosmological constant}	
	{Орбиты массивных частиц в сферически симметричном гравитационном поле с учетом космологической постоянной} 
	{Orbits of massive particles in a spherically symmetric gravitational field in view of cosmological constant} 

\Grant{Работа поддержана Министерством науки и высшего образования Российской Федерации (проект FEUZ20230017).} 

\Abstract%
{\hyphenpenalty=10000
В данной работе представлены результаты теоретического исследования траекторий массивных тел в метрике Коттлера с учетом космологической постоянной $\Lambda$. В работе предложена классификация траекторий в одночастичном случае для метрики с положительной и отрицательной космологической постоянной, перебор вариантов основан на различных решениях уравнения траектории, полученных разложением соответствующей алгебраической кривой в ряд Пюизе. В работе также освещены некоторые представляющие интерес типы траекторий, связанные с различными значениями космологической постоянной. Для случая отрицательной космологической постоянной получена ее верхняя оценка на основании анализа кривых вращения галактик. 

}%
{\hyphenpenalty=10000
In this paper we present the results of a theoretical study of the trajectories of massive particles in the K{\"o}ttler metric in view of the cosmological constant $\Lambda$. For both negative and positive signs of $\Lambda$ a classification of trajectories is proposed, with entries based on different solutions of the trajectory equation, obtained by the expansion of the corresponding algebraic curve in Puiseux series. We also provide some specific types of trajectories which correspond to different values of the cosmological constant. In the case of negative values of the cosmological constant its upper limit is estimated from the galaxy rotation curves. 
	
}

\Key%
  {Общая теория относительности, космологическая постоянная, орбиты}
  {General Relativity, cosmological constant, orbits}


\Author%
{Р.\,С.~Накибов} 
{\textbf{Накибов Руслан Субхиддинович}, магистр физики, ассистент департамента фундаментальной и прикладной физики, ФГАОУ ВО Уральский федеральный университет, ул. Мира, д. 19, г. Екатеринбург, 620002, Россия.}
{nakibov.ruslan@urfu.ru} 
{R.\,S.~Nakibov}
{\textbf{Nakibov Ruslan Subkhiddinovich}, M.Phys., assistant at the department of fundamental and applied physics, Ural Federal University, Mira st., 19, Yekaterinburg, 620002, Russia.}
{a} 
{1} 

\Author%
{А.\,В.~Урсулов}
{\textbf{Урсулов Андрей Владимирович}, к.ф.-м.н., доцент департамента фундаментальной и прикладной физики,  Уральский федеральный университет, ул. Мира, д. 19, г. Екатеринбург, 620002, Россия.}
{AV.Ursulov@urfu.ru} 
{A.\,V.~Ursulov} 
{\textbf{Ursulov Andrey Vladimirovich}, Ph.D., Associate Professor at the department of fundamental and applied physics, Ural Federal University, Mira st., 19, Yekaterinburg, 620002, Russia.}
{a} 
{2} 

\affili {a} 
				{Уральский федеральный университет, г. Екатеринбург, 620002, Россия.}
				{Ural Federal University, Yekaterinburg, 620002, Russia.}

\DOI {00.000000/issn2226-8812.0000.0.0-00}

\newcommand{\pX}{{\mathcal X}}
\let\msf=\mathsf
\newcommand{\var}{\mathop{\sf Var}}

	\Header
	\captionsenglish
	\SubHeader

\section*{Introduction}

It was firstly pointed in 1916 by Einstein that the $\Lambda$-term (cosmological constant) should be included in gravitational field equations within the static Universe model\cite{Einstein}. Later in 1918 K{\"o}ttler considered $\Lambda$-term in a centrally-symmetric gravitational field \cite{Kottler}. Subsequently, the study of the cosmological constant faded into the background, since, as it was believed, there were not enough physical grounds to take it into account\cite{LL}. Nowadays there is a renewed interest in exploring the contribution of the Lambda-term in the Einstein-Hilbert field equations. On the one hand, in the framework of $\Lambda$CDM model the presence of the $\Lambda$--term explains the observed accelerating expansion of the Universe\cite{qq}. On the other hand, diverting the value of cosmological  constant from established in $\Lambda$CDM leads to explanation of some other observed astrophysical phenomena (e.g. rotational curves of some galaxies\cite{Aryal,Farnes}). Besides, the presence of cosmological constant has an effect on properties of black holes and event horizon\cite{Zakharov,Cruz}, on deflection of light and gravitational lensing\cite{Sim,Lake}, and on orbits of massive particles in centrally-symmetric gravitational field\cite{Ono,Solanki}. The study of gravitational fields with negative $\Lambda$ \cite{Aryal,Chrusciel} that induce the anti-de Sitter spacetimes\cite{Gibbons} is particularly interesting.

\section{General statements}
Consider a metric tensor of centrally-symmetric gravitational field in coordinates $x^i=\{ct,r,\theta,\varphi\}$ of a remote observer:
\begin{equation}
\label{mainaxes}
g_{ik}=\left(
\begin{array}{cccc}
\alpha^2 & 0 & 0 & 0\\
0 & -\beta^2 & 0 & 0\\
0 & 0 & -\gamma^2 & 0\\
0 & 0 & 0 & -\gamma^2 \sin^2 \theta\\
\end{array}
\right),
\end{equation}
where $\alpha,\beta,\gamma$ are continuous functions with variable  $r$,  $\theta$ is an angle\cite{LL}. Without loss of generality consider a motion in an equatorial plane $\theta=\pi/2$. The Hamilton-Jacobi equations of a particle of mass $m$ are given~by:
\begin{equation}
\label{HJ}
g^{ik}\frac{\partial S}{\partial x^{i}}\frac{\partial S}{\partial x^{k}}-m^2c^2=0
\end{equation}
where $S$ is action, $E$ is energy, $l$ is an angular momentum of a massive particle and $c$ is the speed of light. 

Substituting
\begin{equation}
\label{HJ ansatz}
S=-Et+l\varphi+S_{r}(r)
\end{equation}
we obtain the equation of trajectory by
\begin{gather}
\label{phi from action}
\frac{\partial S}{\partial l}=\varphi_{0}=const\\
\label{gen eq g_ij}
\varphi-\varphi_0=j\int\frac{dr}{\sqrt{-\frac{\gamma^2}{\beta^2}\left(\gamma^2+j^2\right)+k^2\frac{\gamma^4}{\alpha^2\beta^2}}}
\end{gather}
where $k=E/mc^2$, $j=l/mc^2$ is a reduced angular momentum. For a centrally-symmetric space $k$ and $j$ are constant in time. The latter can also be interpreted as a characteristic radius dependant on mass and angular momentum since it is measured in meters. 

We obtain the potential energy $U(u)$ from the Binet equation\cite{gld},
\begin{eqnarray}
\label{Bine}
\frac{d^2 u}{d \varphi^2}+u=-\frac{1}{mc^2j^2}\frac{dU(u)}{du},\\
U(r)=-\frac{1}{2}mc^2 \left(j^2u^2+u^4F(r)\right).
\end{eqnarray}
where $u=1/r$. Effective potential of a two body problem is given as
\begin{equation}\label{effective-potential}
    U_{eff}(r)=-\frac{mc^2}{2}\frac{F(r)}{r^4}
\end{equation}
Here $F(r)$ is a radicand from \eqref{gen eq g_ij}:
\begin{equation}
    F(r)=-\frac{\gamma^2}{\beta^2}\left(\gamma^2+j^2\right)+k^2\frac{\gamma^4}{\alpha^2\beta^2}
\end{equation}

For the K{\"o}ttler metric with a nonzero cosmological constant $\Lambda$ and a central body of Schwarzschild radius~$r_g$
\begin{equation}
\label{Kottler}
\alpha^2=1-\frac{r_g}{r}-\frac{\Lambda r^2}{3},\quad
\beta^2=\frac{1}{\alpha^2},\quad
\gamma^2=r^2,
\end{equation}
the function $F(r)$ takes the following form
\begin{equation}\label{F(r), Kottler}
F(r)=\frac{\Lambda}{3}r^6+Kr^4+r_g r^3-j^2 r^2+j^2r_g r,
\end{equation}
where $K=-(1-k^2)-\frac{1}{3}\Lambda j^2$.

In the following calculations we assume $r>r_g$, where $r_g$ is an event horizon defined by the equation
\begin{equation}\label{eh}
    1-\frac{r_g}{r}-\Lambda r^2 = 0.
\end{equation} 

\section{Rotatonal curves}
Consider $\Lambda<0$. In this case, it becomes possible to explain the problem of galaxy rotational curves\cite{Aryal}. In the classical limit the linear velocity of a circular motion is given by
\begin{equation}
\label{v1}
v=\sqrt{\frac{1}{m}r\frac{dU}{dr}}
\end{equation}

Assuming $j=r v(r)/c$ we obtain a following rotational curve
\begin{equation}\label{rotation}
    \frac{v}{c}=\sqrt{\frac{\frac{1}{3}|\Lambda| r^2+\frac{1}{2}\frac{r_g}{r}}{2-\frac{3}{2}\frac{r_g}{r}}}.
\end{equation}

Over large distances the rotational curve given by \eqref{rotation} expresses a slow linear growth $\sim \! \sqrt{|\Lambda|}r$. The linear speed has a minimum value if the inequality $r_g^2<|\Lambda|^{-1}$ holds, which is the case for a motion outside the event horizon. The radius $r_{min}$ at which the minimum speed is reached and the corresponding speed value $v_{min}$ are given by
\begin{eqnarray}
    r_{min}=\sqrt[3]{\frac{3}{4}\frac{r_g}{|\Lambda|}}+\frac{3}{8}r_g\\
    v_{min}=\frac{c}{2}\sqrt[6]{\frac{9}{2}|\Lambda| r_g^2}.
\end{eqnarray}

Since the linear speed exceeds its minimum $v\geqslant v_{min}$ a constraint can be placed on the value of $\Lambda$ observed in galaxies.
\begin{equation}
\label{abs Lambda}
|\Lambda|\leqslant\frac{2}{9}\left(2\frac{v_{min}}{c}\right)^6\frac{1}{r_g^2}=8\left(\frac{2}{3}\right)^2\frac{1}{G^2c^4}\frac{v_{min}^6}{M^2}\approx 10^{-13}\frac{v_{min}^6}{M^2}
\end{equation}
where $G$ is a gravitational constant, $M$ is a central body mass and $\Lambda<0$. For a supermassive black hole of mass  $M=10^5 M_\odot$ and $v_{min}\approx200$ km/s one gets $|\Lambda|<10^{-52}$ m$^{-2}$, which is consistent with estimations given in \cite{Aryal}.

\section{Circular motion}
In the classical limit stable circular orbits correspond to minima of effective potential energy  \eqref{effective-potential}, unstable orbits to maxima. Assuming \eqref{F(r), Kottler} and $\Lambda<0$ we get the effective potential 
\begin{equation}
    U_{eff}(r)=-\frac{mc^2}{2}\frac{1}{r^4}\left(-\frac{1}{3}|\Lambda| r^6+ K r^4+r_gr^3-j^2r^2+r_gj^2r\right)
\end{equation}
Finding the extrema of effective potential leads to an equation
\begin{equation}
    \frac{1}{3}|\Lambda| r^5+\frac{1}{2}r_g r^2-j^2r+\frac{3}{2}r_gj^2=0,
\end{equation}
which was solved by the Puiseux method (p. \pageref{pui}), the roots of the equation are presented in the Table \ref{circular roots}.
\begin{table}[h]
    \caption{\label{circular roots}The radii of circular roots in the case of negative cosmological constant. The first case recreates Schwarzschild circular orbits, the second case is a novelty and arises if inequality $J>J_c$ holds, where $J=j/r_g$.}
\setlength\tabcolsep{4.5pt}
\begin{tabular}{|c|c|c|c|l|}\hline
No.&    Conditions                                                   & Unstable root               & Stable root                                   & In dimensionless $\rho=r_g/r$                      \\ \hline
 1&   $j<r_g^{2/3}/|\Lambda|^{1/6}, j>r_g, r_g^2>|\Lambda|j^2$     & $r_1\approx{3}r_g/2$        & $r_2\approx 2{j^2}/{r_g}$                     & $\rho_1={3}/{2},  \rho_2=2J^2$                \\ 
 2&   $j>r_g^{2/3}/|\Lambda|^{1/6}, j>r_g, j^2>|\Lambda|r_g^4$   & $r_1\approx{3}r_g/{2}$      & $r_2\approx\sqrt[4]{{3j^2}/{|\Lambda|}}$      & $\rho_1={3}/{2},  \rho_2=\sqrt{J} J_c^{3/2}$  \\ \hline
\end{tabular}
\end{table}

\begin{figure}[h]
    \centering
    \includegraphics[scale=0.4]{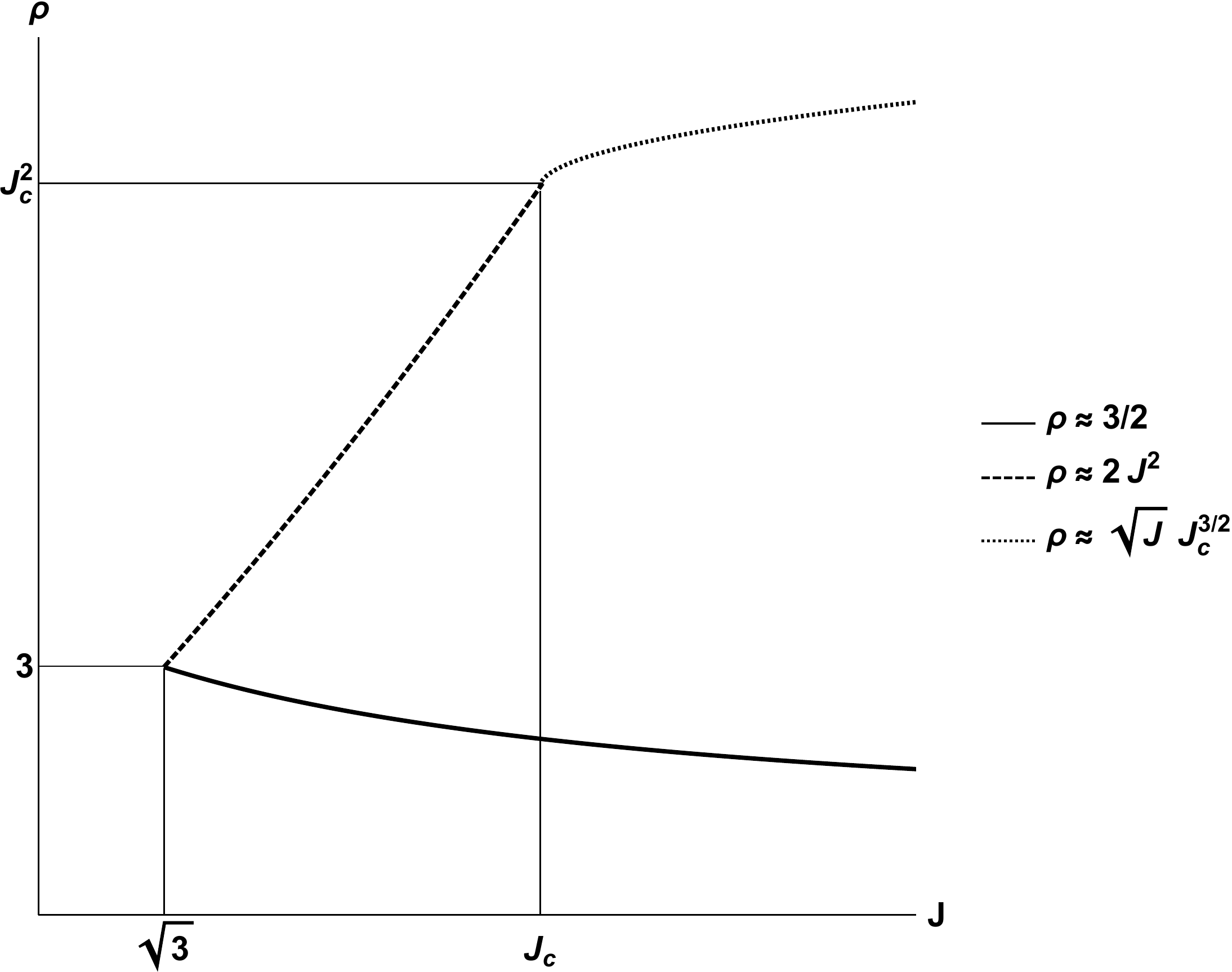}
    \caption{\label{1}Stability diagram for circular orbits in dimensionless coordinates. The upper branch corresponds to stable roots, the lower branch to unstable.}
    \label{fig:enter-label}
\end{figure}

Concluding, for particles with angular momentum high enough or mass low enough so that the value of a parameter~$J$ exceeds the critical value and the sign changes in the inequality $J<J_c$, where 
\begin{equation}
    J_c=\sqrt[6]{3/(|\Lambda| r_g^2)},
\end{equation} 
the stable Schwarzschild root changes to a new value that is expressed explicitly through the cosmological constant. For such particles the gravitational pull is mostly provided by the negative cosmological constant. Fig. \ref{1} depicts the stability diagram for the orbits.

\section{Non-circular motion. Classification and example orbits}

All possible trajectories of non-circular orbits are determined by the equation \eqref{phi from action} where $F(r)$ is provided by \eqref{F(r), Kottler}. Trajectories depend on the number and value of positive roots of the  $p(r)=F(r)/r,$ $(r\neq0)$. 
\begin{equation}\label{puiex}
    p(r)=-\frac{1}{3}|\Lambda|r^5+Kr^3+r_g r^2-j^2 r+j^2r_g=0
\end{equation}
It is possible to get some information by applying the Descartes' rule of signs\cite{descart}. For $\Lambda<0$ there exist three roots or a singular root, for $\Lambda>0$ there are four, two or zero positive roots. 

The equation was solved and the roots are present in the Table \ref{noncircular roots}. The roots are divided in eight configurations, and configurations $A$, $B$, and $C$ represent cases where finite motion is possible. Since the motion is possible if $p(r)>0$, then in configurations with a single root ($r<r_1$) the particle may fall inside the central body or the system is located within event horizon, in double root configurations unbound trajectories are found outside the larger radius ($r>r_2$). In a three or five root configurations there exists a region of finite motion ($r_2<r<r_3$). 

Note that in the configuration $C$ with $K<0$ the lambda-term can be neglected, thus reproducing the Schwarzschild solution. In other cases, the larger root depends on the cosmological constant, only these cases will be considered further.

\begin{table}[h!]
    \caption{\label{noncircular roots}List of configurations and conditions that lead to such configurations. The order of roots can always be determined via conditions; $\sqrt[n]{1}$ is evaluated in a complex plane; only the first term of roots is present.}
\setlength\tabcolsep{9pt}    
\begin{tabular}{|c|c|c|c|c|c|}
\hline
  No. &
  Conditions &
  \multicolumn{4}{c|}{Roots} \\ \cline{3-6}
 &
   &
  $\Lambda>0,\, K>0$ &
  $\Lambda>0,\, K<0$ &
  $\Lambda<0,\, K<0$ &
  $\Lambda<0,\, K>0$ \\ \hline
1 &
  \begin{tabular}[c]{@{}l@{}}$|\Lambda|^3r_g^2j^4>|K|^5$, \\ $j^3|\Lambda|>r_g,$\\$|\Lambda|r_g^4>j^2$\\\end{tabular} &
  $\sqrt[5]{-1}\sqrt[5]{\frac{3j^2r_g}{\Lambda}}$ &
  $\sqrt[5]{-1}\sqrt[5]{\frac{3j^2r_g}{\Lambda}}$ &
  $\sqrt[5]{1}\sqrt[5]{\frac{3j^2r_g}{|\Lambda|}}$ &
  $\sqrt[5]{1}\sqrt[5]{\frac{3j^2r_g}{|\Lambda|}}$\\ \hline
2 &
  \begin{tabular}[c]{@{}l@{}}$|K|^3>|\Lambda|r_g^2$,\\ $|K|^2>|\Lambda|j^2$,\\$|K|^5>|\Lambda|^3j^4r_g^2$,\\$|K|j>r_g$,\\$\sqrt{|K|}r_g>j$,\\\end{tabular} &
  \setlength\tabcolsep{1.5pt}\begin{tabular}[c]{@{}l@{}}$\pm i\sqrt{\frac{K}{\Lambda}}$,\\$\sqrt[3]{-1}\sqrt[3]{\frac{j^2r_g}{K}}$\end{tabular} &
  \setlength\tabcolsep{1.5pt}\begin{tabular}[c]{@{}l@{}}$\pm 1\sqrt{\frac{|K|}{\Lambda}},$\\$\sqrt[3]{1}\sqrt[3]{\frac{j^2r_g}{|K|}}$\end{tabular} &
  \setlength\tabcolsep{1.5pt}\begin{tabular}[c]{@{}l@{}}$\pm 1\sqrt{\frac{K}{|\Lambda|}},$\\$\sqrt[3]{-1}\sqrt[3]{\frac{j^2r_g}{K}}$\end{tabular} &
  \setlength\tabcolsep{1.5pt}\begin{tabular}[c]{@{}l@{}}$\pm i\sqrt{\frac{|K|}{|\Lambda|}},$\\$\sqrt[3]{1}\sqrt[3]{\frac{j^2r_g}{|K|}}$\end{tabular}  \\ \hline
3 &
  \begin{tabular}[c]{@{}l@{}}$|\Lambda|r_g^2>|K|^3$,\\$r_g^4>|\Lambda|j^6,$\\$r_g>|\Lambda|j^3,$\\$r_g>j,$\\\end{tabular}&
  \begin{tabular}[c]{@{}l@{}}$\sqrt[3]{-1}\sqrt[3]{\frac{r_g}{\Lambda}},$\,$\pm ij$ \end{tabular}&
  \begin{tabular}[c]{@{}l@{}}$\sqrt[3]{-1}\sqrt[3]{\frac{r_g}{\Lambda}},$\,$\pm ij$ \end{tabular}&
  \begin{tabular}[c]{@{}l@{}}$\sqrt[3]{1}\sqrt[3]{\frac{r_g}{|\Lambda}|},$\,$\pm ij$ \end{tabular}&
  \begin{tabular}[c]{@{}l@{}}$\sqrt[3]{1}\sqrt[3]{\frac{r_g}{|\Lambda}|},$\,$\pm ij$\end{tabular}\\ \hline
4 &
  \begin{tabular}[c]{@{}l@{}}$|\Lambda|j^2>K^2,$\\$j^6|\Lambda|>r_g^4,$\\$j^2>|\Lambda|r_g^4,$\\\end{tabular}&
  \begin{tabular}[c]{@{}l@{}}$\sqrt[4]{1}\sqrt[4]{\frac{j^2}{\Lambda}},$\,$r_g$ \end{tabular}&
  \begin{tabular}[c]{@{}l@{}}$\sqrt[4]{1}\sqrt[4]{\frac{j^2}{\Lambda}},$\,$r_g$ \end{tabular}&
  \begin{tabular}[c]{@{}l@{}}$\sqrt[4]{-1}\sqrt[4]{\frac{j^2}{|\Lambda|}},$\,$r_g$ \end{tabular}&
  \begin{tabular}[c]{@{}l@{}}$\sqrt[4]{-1}\sqrt[4]{\frac{j^2}{|\Lambda|}},$\,$r_g$\end{tabular}\\\hline
5 &
  \begin{tabular}[c]{@{}l@{}}$|K|^3>|\Lambda|r_g^2,$\\$|K|^2>|\Lambda|j^2,$\\$|K|^5>|\Lambda|^3j^4r_g^2,$\\$r_g^2>j^2|K|,$\\$r_g>j|K|,$\\$r_g>j,$\\\end{tabular}&
  \begin{tabular}[c]{@{}l@{}}$\pm i\sqrt{\frac{K}{\Lambda}},$\\$-\frac{r_g}{K},$\,$\pm ij$ \end{tabular}&
  \begin{tabular}[c]{@{}l@{}}$\pm 1\sqrt{\frac{|K|}{\Lambda}},$\\$\frac{r_g}{|K|},$\,$\pm ij$ \end{tabular}&
  \begin{tabular}[c]{@{}l@{}}$\pm 1\sqrt{\frac{K}{|\Lambda|}},$\\$-\frac{r_g}{K},$\,$\pm ij$ \end{tabular}&
  \begin{tabular}[c]{@{}l@{}}$\pm i\sqrt{\frac{|K|}{|\Lambda|}},$\\$\frac{r_g}{|K|},$\,$\pm ij$\end{tabular}\\ \hline
A(6) &
  \begin{tabular}[c]{@{}l@{}}$|\Lambda|r_g^2>|K|^3,$\\$r_g^4>|\Lambda|j^6,$\\$r_g>|\Lambda|j^3,$\\$j>r_g$\\\end{tabular}&
  \begin{tabular}[c]{@{}l@{}}$\sqrt[3]{-1}\sqrt[3]{\frac{r_g}{\Lambda}},$\\$\frac{j^2}{r_g},$\,$r_g$ \end{tabular}&
  \begin{tabular}[c]{@{}l@{}}$\sqrt[3]{-1}\sqrt[3]{\frac{r_g}{\Lambda}},$\\$\frac{j^2}{r_g},$\,$r_g$ \end{tabular}&
  \begin{tabular}[c]{@{}l@{}}$\sqrt[3]{1}\sqrt[3]{\frac{r_g}{|\Lambda}|},$\\$\frac{j^2}{r_g},$\,$r_g$ \end{tabular}&
  \begin{tabular}[c]{@{}l@{}}$\sqrt[3]{1}\sqrt[3]{\frac{r_g}{|\Lambda}|},$\\$\frac{j^2}{r_g},$\,$r_g$\end{tabular}\\ \hline
B(7) &
  \begin{tabular}[c]{@{}l@{}}$K^3>|\Lambda|r_g^2,$\\$K^2>|\Lambda|j^2,$\\$K^5>|\Lambda|^3j^4r_g^2,$\\$j^2K>r_g,$\\$j^2>Kr_g,$\\\end{tabular}&
  \begin{tabular}[c]{@{}l@{}}$\pm i\sqrt{\frac{K}{\Lambda}},$\\$\pm 1\frac{j}{\sqrt{K}},$\,$r_g$ \end{tabular}&
  \begin{tabular}[c]{@{}l@{}}$\pm 1\sqrt{\frac{|K|}{\Lambda}},$\\$\pm i\frac{j}{\sqrt{K}},$\,$r_g$ \end{tabular}&
  \begin{tabular}[c]{@{}l@{}}$\pm 1\sqrt{\frac{K}{|\Lambda|}},$\\$\pm 1\frac{j}{\sqrt{K}},$\,$r_g$ \end{tabular}&
  \begin{tabular}[c]{@{}l@{}}$\pm i\sqrt{\frac{|K|}{|\Lambda|}},$\\$\pm i\frac{j}{\sqrt{K}},$\,$r_g$\end{tabular}\\ \hline
C(8) &
  \begin{tabular}[c]{@{}l@{}}$K^3>|\Lambda|r_g^2,$\\$K^2>|\Lambda|j^2,$\\$|K|^5>|\Lambda|^3j^4r_g^2,$\\$r_g^2>j^2|K|,$\\$r_g>j|K|,$\\$j>r_g$\end{tabular} &
  \begin{tabular}[c]{@{}l@{}}$\pm i\sqrt{\frac{K}{\Lambda}},$\\$-\frac{r_g}{K},$\,$\frac{j^2}{r_g},$\,$r_g$ \end{tabular}&
  \begin{tabular}[c]{@{}l@{}}$\pm 1\sqrt{\frac{|K|}{\Lambda}},$\\$\frac{r_g}{|K|},$\,$\frac{j^2}{r_g},$\,$r_g$ \end{tabular}&
  \begin{tabular}[c]{@{}l@{}}$\pm 1\sqrt{\frac{K}{|\Lambda|}},$\\$-\frac{r_g}{K},$\,$\frac{j^2}{r_g},$\,$r_g$ \end{tabular}&
  \begin{tabular}[c]{@{}l@{}}$\pm i\sqrt{\frac{|K|}{|\Lambda|}},$\\$\frac{r_g}{|K|},$\,$\frac{j^2}{r_g},$\,$r_g$\end{tabular}\\ \hline
\end{tabular}
\end{table}

\begin{figure}[h]
\begin{minipage}{0.47\linewidth}
    \centering
    \includegraphics[scale=0.3]{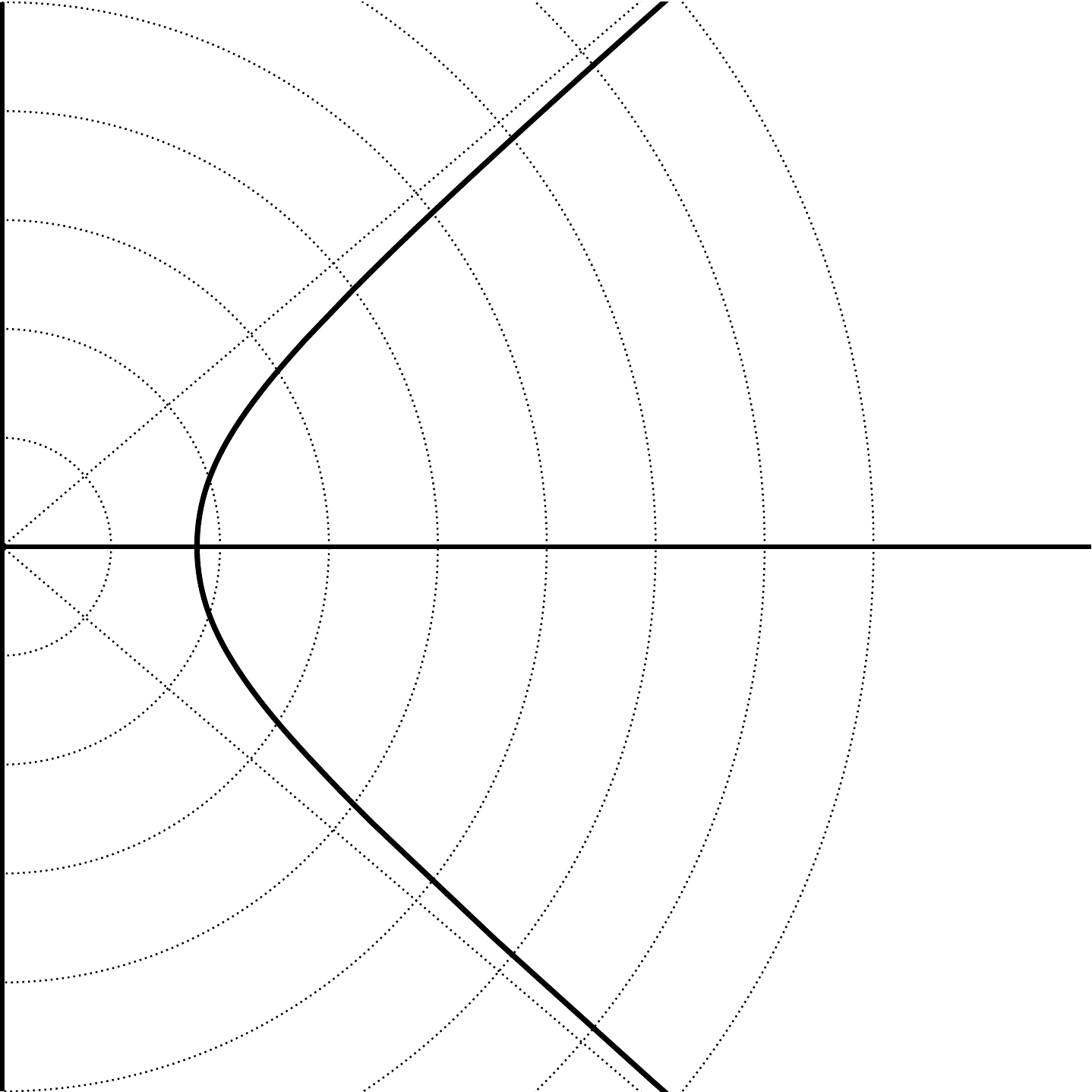}
    \caption{\label{2}Unbound orbit with $\Lambda>0$, $K>0$}
\end{minipage}\hfill%
\begin{minipage}{0.47\linewidth}
    \centering
    \includegraphics[scale=0.3]{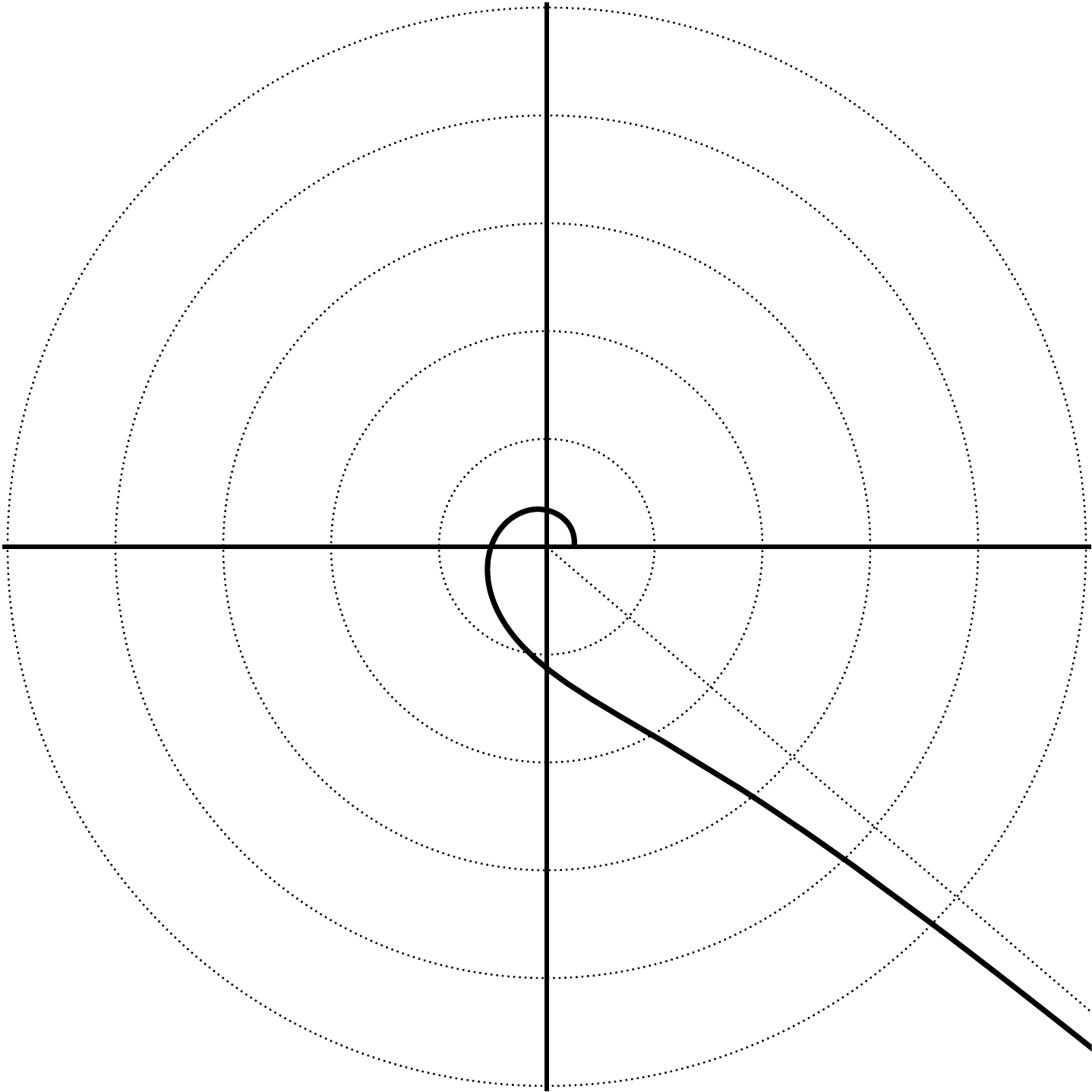}
    \caption{\label{3}Unbound orbit with $\Lambda>0$, $K<0$}
\end{minipage}\hfill%
\end{figure}

Let us list some of the orbits. Unbound trajectories outside $r_{max}=\sqrt{3|K|/\Lambda}$ in case of $\Lambda>0$ are either modified hyperbolic spirals if $K>0$, or modified hyperbolas if $K<0$: 
\begin{equation}
    r=\frac{1}{\sqrt{\frac{\Lambda}{3K}-\frac{K\varphi^2}{j^2}}}
\end{equation}
The formula covers both cases. Corresponding trajectories are displayed in Fig. \ref{2} and Fig. \ref{3}

Consider a case of three roots and $\Lambda<0$. Let us factorize the radical expression of \eqref{phi from action}:
\begin{equation}\label{int-roots2}
    \varphi=\int\frac{jdr}{\sqrt{-\frac{1}{3}|\Lambda|r(r-r_1)(r-r_2)(r-r_3)(r-r_4)(r-r_5)}}
\end{equation}
In each of configurations $A$,\,$B$,\,$C$ (see Table \ref{noncircular roots}) there is a total of five roots, some of which may be negative or complex. The root $r=r_g$ is present only once in each of those configurations and it's the smallest positive root. Assuming that the orbits take place in $r_{max}>r>r_{min}\gg r_g$, we can expand the factor $(r-r_g)^{-1/2}$ as
\begin{equation}
    \frac{1}{\sqrt{r-r_g}}\approx\frac{1}{\sqrt{r}}+\frac{1}{2}\frac{r_g}{r\sqrt{r}}+\dots
\end{equation}
in which case trajectories can be expressed in a form of a linear combination of incomplete elliptic integrals and elementary functions.
\begin{equation}\label{el int}
    \varphi=\sqrt{\frac{3j^2}{|\Lambda|}}\left[\int\frac{dr}{r\sqrt{(r_1-r)(r-r_2)(r-r_3)(r-r_4)}}+\frac{r_g}{2}\int\frac{dr}{r^2\sqrt{(r_1-r)(r-r_2)(r-r_3)(r-r_4)}}\right]
\end{equation}
The corresponding formulas for the case of four real roots (conf.~$B,\,C$) and two real and two complex roots (conf.~$A$) are taken from\cite{elli}. Here $F(\vartheta|m),\,E(\vartheta|m),\, \Pi(\alpha^2;\vartheta|m)$ are incomplete elliptic integrals of first, second and third kind. Let us introduce the notation
\begin{gather*}
    A^2=(a-\text{Re}(c))^2+\text{Im}(c)^2, \quad B^2=(b-\text{Re}(c))^2+\text{Im}(c)^2, \quad g=1/\sqrt{AB}\\
    \alpha=\frac{bA-aB}{aB+bA},\quad \alpha_1=\frac{A-B}{A+B},
    \quad X_1=1,\quad X_2=\frac{r_g}{2}  \\
    m=\frac{(a-b)^2-(A-B)^2}{4AB},\quad\theta=\arccos\frac{(a-r)B-(r-b)A}{(a-r)B+(r-b)A}, \quad \cos\theta=\text{cn}u\\
    R_{-2}=\frac{1}{m}((m-\alpha^2(1-m))F(\theta|m)+\alpha^2E(\theta|m)+2\alpha\sqrt{m}\arccos (\text{dn} u))\\
    R_{-1}=F(\theta|m)+{\frac{\alpha}{\sqrt{m}}}\arccos (\text{dn} u), \quad R_0=F(\theta|m) \\
    R_{1}=\frac{1}{1-\alpha^2}\left(\Pi(\frac{\alpha^2}{\alpha^2-1};\theta|m)-\alpha\sqrt{\frac{1-\alpha^2}{m+(1-m)\alpha^2}}\arctan\left(\sqrt{\frac{m+(1-m)\alpha^2}{1-\alpha^2}}\text{sd}u\right)\right) \\
    R_{2}=\frac{1}{(\alpha^2-1)(m+\alpha^2(1-m))}\left((\alpha^2(2m-1)-2m)R_1+2mR_{-1}-mR_{-2}+\frac{\alpha^3\text{sn}u\,\text{dn}u}{1+\alpha\,\text{cn}u}\right),
\end{gather*}
where $a>b, c, \overline{c}$ are the roots of the radicand. Then for configuration $A$ trajectories are given by the  formula:
\begin{equation}\label{conf6}
    \varphi(r)=\sqrt{\frac{3j^2}{|\Lambda|}}\sum_{s=1}^2\frac{(A+B)^sgX_s}{(Ab-Ba)^s}\sum_{n=0}^s\frac{\alpha_1^{s-n}(\alpha-\alpha_1)^ns!}{(s-n)!n!}R_n
\end{equation}
Introducing the notation
\begin{gather*}
    g=\frac{2}{\sqrt{(a-c)(b-d)}},\quad \alpha=\frac{c}{b}\frac{a-b}{a-c} ,\quad \alpha_1=\frac{b}{c}\alpha,\quad X_1=1,\quad X_2=\frac{r_g}{2}  \\
    m=\frac{(a-b)(c-d)}{(a-c)(b-d)},\quad\theta=\arcsin\sqrt{\frac{(a-c)(r-b)}{(a-b)(r-c)}}\\
    V_0=F(\theta|m),\quad V_1=\Pi(\alpha^2,\theta|m),\quad \sin\theta=\text{sn} u \\
\begin{split}
    V_2=\frac{1}{2(\alpha^2-1)(m-\alpha^2)} (\alpha^2 &E(\theta|m)+(m-\alpha^2)F(\theta|m)+\\+&\left.(2\alpha^2m+2\alpha^2-\alpha^4-3m)\Pi(\alpha^2,\theta|m)-\frac{\alpha^4 \text{sn}u\, \text{cn}u\, \text{dn}u}{1-\alpha^2 \text{sn}^2u}\right),
\end{split}
\end{gather*}
where $a>b>c>d$ are roots of the radicand we get trajectories formula for configurations $B$ and $C$:
\begin{equation}\label{conf8}
    \varphi(r)=\sqrt{\frac{3j^2}{|\Lambda|}}\sum_{s=1}^2\frac{gX_s}{b^s}\frac{\alpha_1^{2s}}{\alpha^{2s}}\sum_{n=0}^s\frac{(\alpha^2-\alpha_1^2)^n s!}{\alpha_1^{2n}n!(s-n)!}V_n
\end{equation}

For any set of parameters $\{\Lambda, r_g, K, j\}$ this process allows to find the set of roots and a suitable formula of trajectory inside the region $r_{max}>r>r_{min}$. These formulas describe the motion from $r_{min}$ to $r_{max}$, then the particle returns along a similar trajectory.

Note two important cases. The first case arises if $\log_{10}(r_{max}/r_{min})>2..3$. The trajectory then is a modified hyperbolic spiral ($\Lambda<0, K>0$):
\begin{equation}
\label{rfi031}
r=\frac{1}{\sqrt{\frac{K\varphi^2}{j^2}+\frac{|\Lambda|}{3K}}}
\end{equation}
The second case describes a closed bound orbit in case of $r_g=0$ and $\Lambda<0$. Integrating the simplified expression 
\begin{equation}
       \varphi=\int\frac{jdr}{r\sqrt{\frac{1}{3}\Lambda r^4+Kr^2-j^2}}.
\end{equation}
we obtain the general formula for such orbits
\begin{equation}
\label{rfi1}
r=\sqrt{\frac{2j^2/K}{1-\sqrt{1+\frac{4}{3}\frac{\Lambda j^2}{K^2}}\sin{(2\zeta\varphi+\Delta\varphi)}}},
\end{equation}
where $\Delta\varphi=0$, $\zeta=1$ in case of $r_g=0$. 

\begin{figure}[h]
\begin{minipage}{0.3\textwidth}
    \includegraphics[scale=0.3]{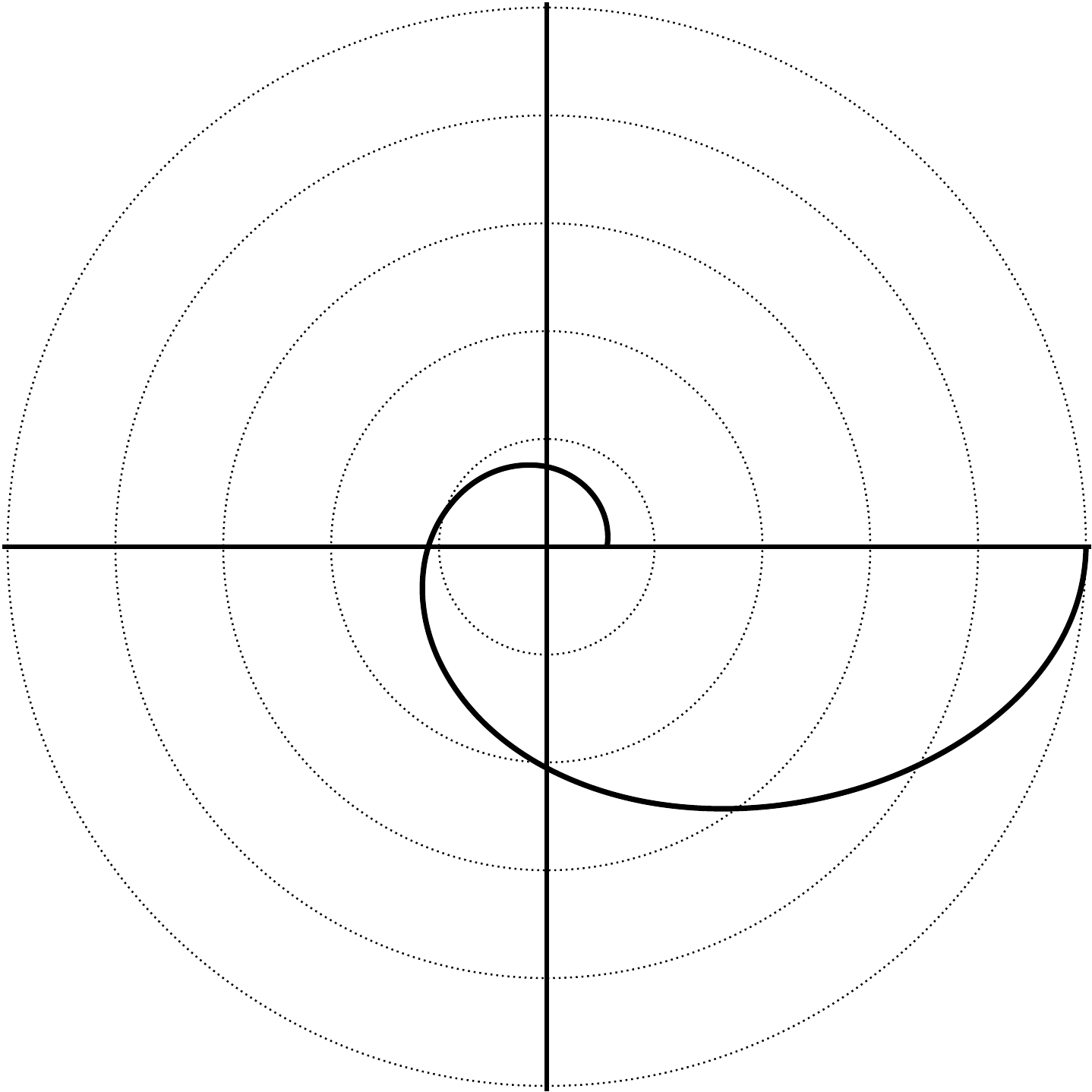}
    \caption{\label{4}Modified hyperbolic spiral  $\Lambda<0$, $K>0$}
\end{minipage}\hfill%
\begin{minipage}{0.3\textwidth}
    \includegraphics[scale=0.3]{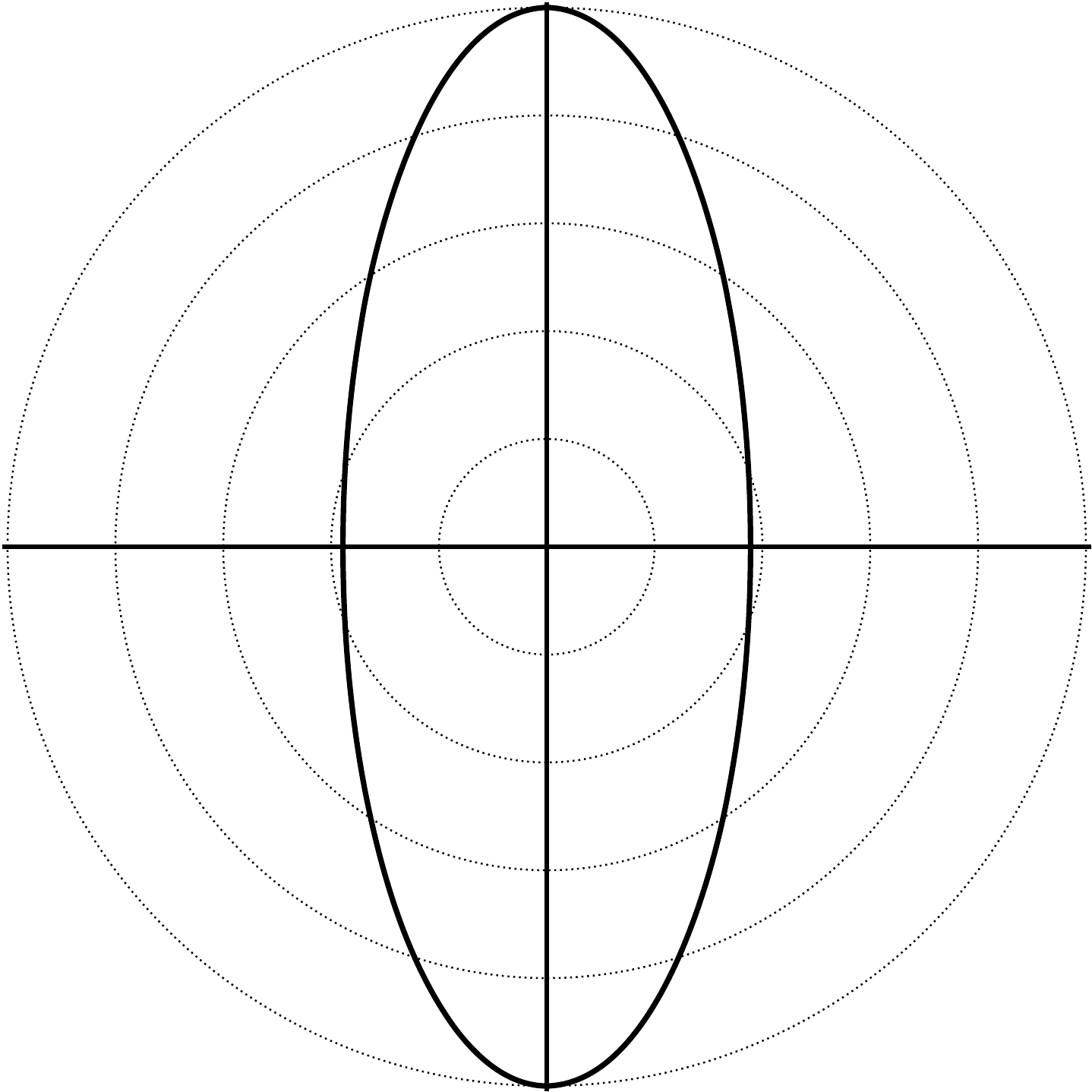}
    \caption{\label{5}Closed orbit with $\Lambda<0$, $K>0$ and $r_g=0$.}
\end{minipage}\hfill%
\begin{minipage}{0.3\textwidth}
    \includegraphics[scale=0.3]{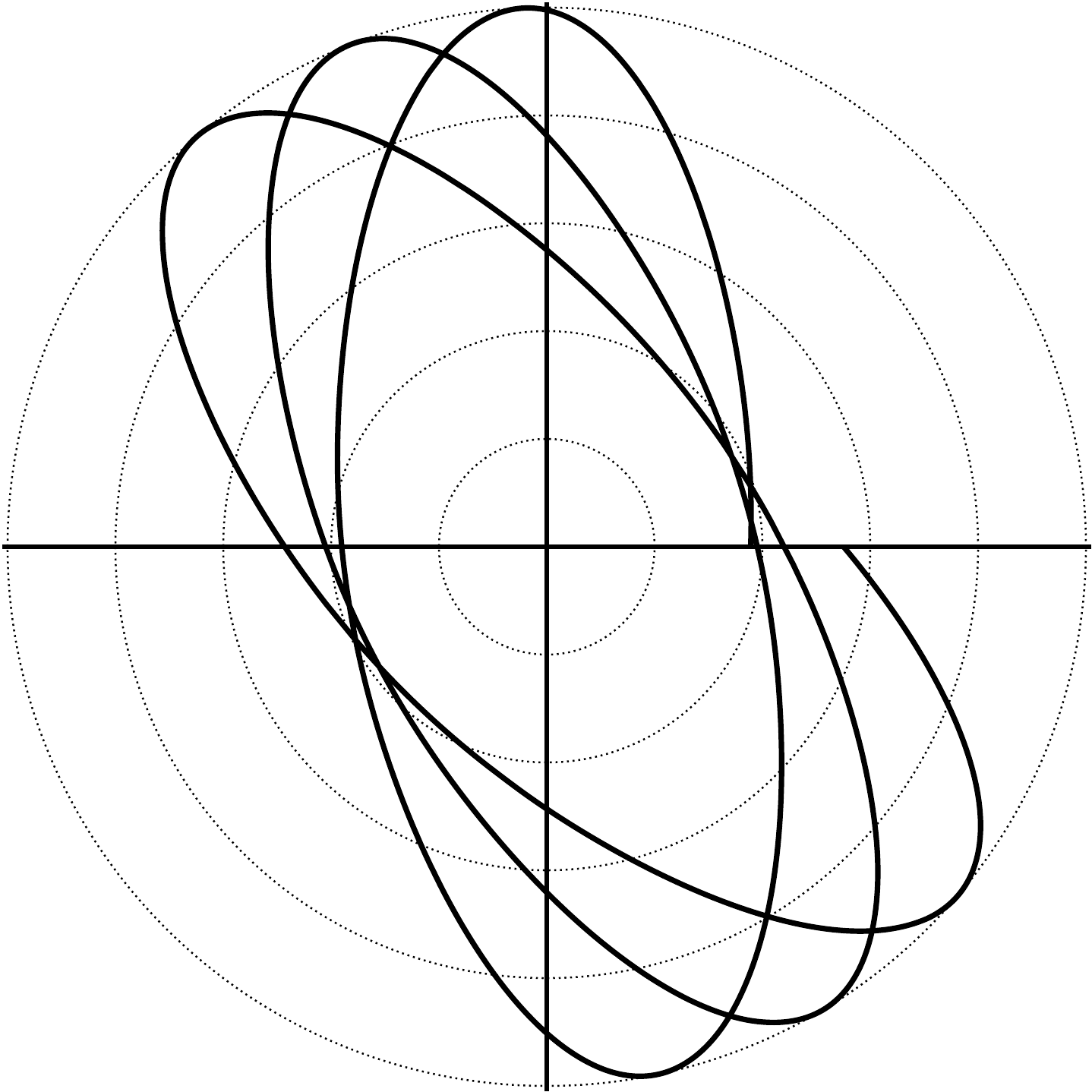}
    \caption{\label{6}Bound orbit with perihelion precession with $\Lambda<0$, $K>0$ due to accounting for $r_g$.}
\end{minipage}\hfill%
\end{figure}

Equation \eqref{rfi1} allows for several trajectories. For $\Lambda>0$, regardless of the sign of $K$, hyperbolic orbits follow from the formula, however, if the attraction to the center is provided only by the cosmological constant $\Lambda<0$, $(\text{and } K>0)$, then the formula implies a closed bound orbit. Accounting for the non-zero gravitational radius leads to deviations in the formula and perihelion precession:
\begin{gather}
    \zeta=r_{min}r_{max} \sqrt{\frac{|\Lambda|}{3j^2}}, \\
    \label{rfi2}\Delta\varphi=-\frac{r_g}{r_{min}}E\left(\arcsin\sqrt{\frac{r_{max}^2(r^2-r_{min}^2)}{r^2(r_{max}^2-r_{min}^2)}}\left|1-\frac{r_{min}^2}{r_{max}^2}\right.\right)
\end{gather}
where $E(\theta|m)$ is an incomplete elliptic integral of the second kind. These variations are displayed in Fig. \ref{4}--\ref{6}. Expressions \eqref{rfi1}--\eqref{rfi2} can be used for configuration $B$ if $r_{min}\gg r_g$ due to both $r_{min},\, r_{max}$ being independent on $r_g$ in the first term. In other configurations radii depend on gravitational radius and formulas  \eqref{conf6} and~\eqref{conf8} should be used. 

\section{A study of trajectories}
\begin{figure}[h]
\centering
    \includegraphics[scale=0.6]{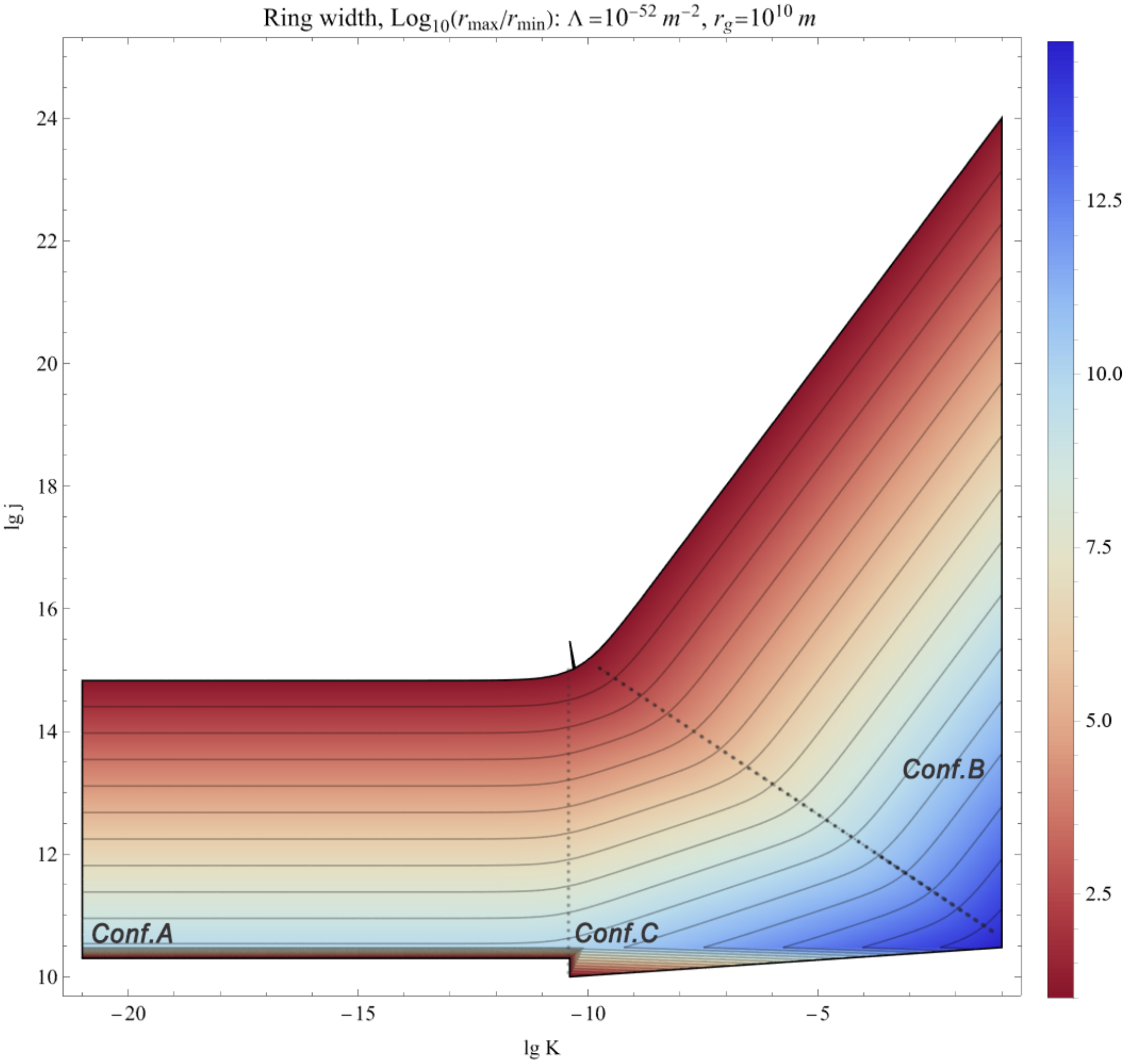}
    \caption{\label{7}An example of a heatmap: for any $j,K$ with fixed $\Lambda,r_g$ a logarithm of the ratio of outer and inner radii is calculated. Larger values correspond to the spiral trajectories, smaller values to bound orbits with precession.}
\end{figure}

Let us fix $r_g$ and $\Lambda$ and vary  parameters $j$ and $K$. Then for each pair $\{j,K\}$ one can calculate the value of any quantity $P$, thus obtaining the dependence $P = P(j, K)$ as a dataset. The function $P = P(j, K)$ then can be represented as a heatmap (e.g. Fig. \ref{7}). Finally, comparing the diagrams for different $r_g$ and $\Lambda$, one can study the influence of all four parameters on the particle trajectories. The script for such procedure is written and can be shared via e-mail.

In general, bound and spiral orbits are mathematically possible with a variety of combinations of parameters, but not all of them could have been observed in physical systems. It is worth noting some interesting properties of orbits in the case of $\Lambda<0$.

Trajectories that belong to narrow regions of motion are located at characteristic radii from 100 pc to 100 kpc, which corresponds to the size of galaxies, and if the system is fairly remote the particles cannot escape. In particular, the orbits are limited by $r_{max}=\sqrt[3]{3r_g/|\Lambda|}$ in configuration $A$ (see Table~\ref{noncircular roots}), which corresponds to low-energy particles. This radius shows weak grows with $r_g$ and also weakly depends on the particle properties ($J$, $K$). 

Trajectories in wider regions of motion form spiral trajectories despite of a bounded region. In this case the maximal radius can approach the size of an observable Universe ($\sqrt{3/\Lambda}$). There is merit in assuming that a particle at such distance would be captured by another gravitating system. However, not all particles can leave the system in a physically reasonable time. For <<slow>> particles, the escape along the spiral trajectory will take a significantly longer time and the particle for the observer will effectively remain on the spiral trajectory.

\section*{Conclusion}
In the present paper we consider the trajectories of massive particles in the K{\"o}ttler metric for various values of the cosmological constant. A classification of these trajectories is proposed, which is based on the Puiseux method for polynomials. Some specific or notable types of trajectories are outlined. It should be noted that not all obtained trajectories can be found in a real system. However, in our opinion, in addition to academic interest, understanding the nature of trajectories may be useful in planning experiments to determine the sign and magnitude of the cosmological constant.

\section*{Appendix. Puiseux method}\label{pui}
In the present paper we use a proposed Puiseux method to analyse the main equation for trajectories. This section outlines some key theoretical points concerning the underlying theory and a proposed application of it as a root-finding method for poynomial equations.   

The Newton-Puiseux theorem was originally devised for the study of algebraic curves. An algebraic curve $F(x,y)=0$ is a polynomial function of each of its variables over the field of complex numbers. The theorem provides a method for finding formulas for all of the branches in the form of generalized power series which allows for negatives and fractions, the Puiseux series\cite{wall_2004}. Given that the highest order of $y$ in $F(x,y)=0$ is $n$, the theorem proves that it is possible to obtain exactly $n$ different expansions $y(x)$ for the branches. For technical details on algorithm see\cite{compute}. 

It is possible to use the Pusieux expansion for the study of polynomials and employ it as a tedious root finding algorithm for polynomials of higher orders. Consider an equation of the form
\begin{equation}\label{j1}
a_n y^n+a_{n-1}y^{n-1}+\dots a_1 y+a_0=0,    
\end{equation}
where $a_{n}$ are parameters (e.g. \eqref{puiex}). It is possible to pick a number $x$ and a set of exponents $b_{i}$ so the following substitution holds
\begin{equation}\label{j}
    a_i=\mathrm{sign} (a_i) x^{b_i}.
\end{equation}
In such form the algebraic equation becomes an algebraic curve and can be expanded in a Puiseux series. The next key point is the usage of Newton polygons. Originally the roots are determined from the slopes of a convex hull of points that represent the terms of a polynomial. Note that different sets of parameters may form the same convex hull, hence yielding same expansions and also note that for given polynomial there is a finite amount of Newton polygons. Consider then a set of all possible Newton polygons. Those are defined by relations of lines and dots which are easy to obtain in the form of sets of inequalities. After inverse substitution these sets become so-called conditions which mark the parameter space of an equation and Puiseux expansions become approximate formulas suitable for those conditions which effectively solves the equation. 

Concluding, we find the main use of the proposed root-finding method not in obtaining precise formulas for equations of higher orders but in extracting valuable information about all possible sets of roots and conditions that induce these roots.


\begin{otherlanguage}{english}


\begin{otherlanguage}{russian}
\Footer
\end{otherlanguage}

\vspace{10pt}
\FooterSub
\end{otherlanguage}


\begin{thebibliography}{99}

\bibitem{Einstein}
Einstein A.
Kosmologische Betrachtungen zur allgemeinen Relativit{\"a}tstheorie.
\textit{Sitzungsberichte der K{\"o}niglich Preussischen Akademie der Wissenschaften}, 1917, pp. 142--152.

\bibitem{Kottler}
K{\"o}ttler F. 
{\"U}ber die physikalischen Grundlagen der Einsteinschen Gravitationstheorie.
\textit{Annalen der Physik}, 1918, vol. 361, pp. 401-462

\bibitem{LL}
Landau L. D., Lifschits E. M.
{\it The Classical Theory of Fields}, Course of Theoretical Physics, Vol.2.
Oxford, Pergamon Press, 1975.

\bibitem{qq}
Planck Collaboration
Planck 2015 results - XIII. Cosmological parameters
\textit{A\&A}, 2016, vol. 594, p. A13. doi: 10.1051/0004-6361/201525830

\bibitem{Aryal}
Aryal B., Pandey R., Baral N., Khanal U., Saurer W.
Estimation of mass and cosmological constant of nearby spiral galaxies using galaxy rotation curve.
2013, arXiv:1307.1824 [astro-ph.CO]

\bibitem{Farnes}
Farnes J. S.
A Unifying Theory of Dark Energy and Dark Matter: Negative Masses and Matter Creation within a Modified $\Lambda$CDM Framework.
\textit{A\&A}, 2016, vol. 620, p. A92. doi: 10.1051/0004-6361/201832898

\bibitem{Zakharov}
Zakharov A. F.
Are signatures of anti-de-Sitter black hole at the Galactic Center?
\textit{arXiv e-prints}, 2014, arXiv:1407.2591. doi: 10.48550/arXiv.1407.2591

\bibitem{Cruz}
Cruz N., Olivarez M., Villanueva J. R.
The golden ratio in Schwarzschild{\textendash}Kottler black holes.
\textit{The European Physical Journal C}, 2017, vol. 77, no. 2. doi: 10.1140/epjc/s10052-017-4670-7

\bibitem{Sim}
Simpson F., Peacock J., Heavens A.
On lensing by a cosmological constant.
\textit{Monthly Notices of the Royal Astronomical Society}, 2010, vol. 402, no. 3, pp. 2009--2016. doi: 10.1111/j.1365-2966.2009.16032.x

\bibitem{Ishak1}
Ishak M., Rindler W., Dossett J.
More on lensing by a cosmological constant.
\textit{Monthly Notices of the Royal Astronomical Society}, 2010, vol. 403, no. 4, pp. 2152-2156. doi: 10.1111/j.1365-2966.2010.16261.x

\bibitem{shak2}
Ishak M., Rindler W.
The relevance of the cosmological constant for lensing.
\textit{General Relativity and Gravitation}, 2010, vol. 42, no. 9, pp. 2247--2268. doi: 10.1007/s10714-010-0973-9

\bibitem{ARAKIDA}
Arakida H., Kasai M.
Effect of the cosmological constant on the bending of light and the cosmological lens equation.
\textit{Physical Review D}, 2012, vol. 85, no. 2. doi: 10.1103/physrevd.85.023006

\bibitem{Lake}
Lake K.
Bending of light and the cosmological constant.
\textit{Physical Review D}, 2002, vol. 65, no. 8 doi: 10.1103/physrevd.65.087301

\bibitem{Ono}
Ono T., Suzuki T., Fushimi N., Yamada K., Asada H.
Application of Sturm's theorem to marginal stable circular orbits of a test body in spherically symmetric and static spacetimes.
 2015, arXiv:1508.00101 [gr-qc]

\bibitem{Solanki}
Solanki R.
Kottler spacetime in isotropic static coordinates.
\textit{Classical and Quantum Gravity}, 2021, vol. 39, no. 1, p. 015015. doi: 10.1088/1361-6382/ac3aa0

\bibitem{Chrusciel}
Chru{\'{s}}ciel P., Simon W.
Towards the classification of static vacuum spacetimes with negative cosmological constant.
\textit{Journal of Mathematical Physics}, 2001, vol. 42, no. 4, pp. 1779--1817. doi: 10.1063/1.1340869

\bibitem{Gibbons}
Gibbons G. W.
Anti-de-Sitter spacetime and its uses.
2011, arXiv:1110.1206 [hep-th]

\bibitem{descart}
Xiaoshen W.
A Simple Proof of Descartes's Rule of Signs.
\textit{The American Mathematical Monthly}, 2004, vol. 111, no. 6, pp. 525-526. doi: 10.1080/00029890.2004.11920108

\bibitem{elli}
Byrd P.F., Friedman M.D.
{\it Handbook of Elliptic Integrals for Engineers and Scientists}. Springler Berlin Heidelberg, 2012.

\bibitem{compute}
Willis N., Didier A., Sonnanburg K.
How to compute Puiseux expansion.
2008, arXiv:0807.4674 [math.AG]

\bibitem{wall_2004}
Wall C. T. C.
{\it Singular Points of Plane Curves}, London Mathematical Society Student Texts. 
Cambridge University Press, 2004. doi: 10.1017/CBO9780511617560

\bibitem{gld}
Goldstein H. 
{\it Classical Mechanics}. Addison-Wesley Publishing Company, 1980.



\end{thebibliography}
\end{document}